\documentclass[referee]{aa}
\usepackage{graphicx}
\newcommand{\be}{\begin{equation}}
\newcommand{\ee}{\end{equation}}
\newcommand{\ls}{LS~5039}
\begin{document}
\title{A numerical model for the $\gamma$-ray emission of the
microquasar \ls}
\author{V. Bosch-Ramon, J. M. Paredes}
\institute{Departament d'Astronomia i
Meteorologia, Universitat de Barcelona, Av. Diagonal
647, 08028 Barcelona, Spain\\\email{vbosch@am.ub.es;
jmparedes@ub.edu}}
\authorrunning{Bosch-Ramon \& Paredes}
\titlerunning{$\gamma$-ray emission in microquasars}
\offprints{V. Bosch-Ramon \\ }
\abstract{The possible association between the microquasar \ls\ and the EGRET source
3EG~J1824$-$1514 suggests that microquasars could also be sources of high-energy $\gamma$-rays. In
this paper, we explore, with a detailed numerical model, if this 
system can produce the emission detected by EGRET ($>$100~MeV) through inverse Compton (IC)
scattering. Our numerical approach considers a population of relativistic electrons entrained 
in a cylindrical inhomogeneous jet, interacting with both the radiation and the magnetic
fields, taking into account the Thomson and Klein-Nishina regimes of interaction. The 
computed spectrum reproduces the observed spectral characteristics at very high energy.
}  

\maketitle

\section{Introduction} \label{intro}

Microquasars are a selected class of X-ray binaries that produce relativistic radio jets
(Mirabel \& Rodr{\'{\i}}guez \cite{Mirabel&rodriguez99}; Fender \cite{Fender04}). The origin
of the jets is related to the matter accreted by the compact object, a neutron star or a
black hole, from the companion star. These systems behave as scaled-down versions of quasars
and active galactic nuclei. The population of microquasars is still a very reduced one, with
about sixteen known objects up to now (Rib\'o \cite{Ribo02t}). Several models have
been developed to explore the high energy emission from the jets of microquasars. Two kinds of model
can be found in the literature depending on whether hadronic or leptonic jet matter dominates the
emission at such an energy range: the hadronic jet models (e.g. Romero et~al.
\cite{Romero03}), and the leptonic jet models. Among leptonic jet models, there are IC jet emission
models that can produce X-rays and $\gamma$-rays, based in some cases on the synchrotron self-Compton
(SSC) process (i.e. Band \& Grindlay \cite{Band&grindlay86}; Atoyan \& Aharonian 
\cite{Atoyan&aharonian99}), and in other cases on external sources for the IC seed photons
(EC) (i.e. Kaufman Bernad\'o et~al. \cite{Kaufman-Bernado02}; Georganopoulos et~al.
\cite{Georganopoulos02}). In addition, there are synchrotron jet emission models that can produce
X-rays (i.e. Markoff et~al. \cite{Markoff01}).

The microquasar nature of LS~5039 was clearly established when non-thermal radiation produced in a
jet was detected (Paredes et~al. \cite{Paredes00}). Interest in this source has grown significantly
because it has turned out to be a source of persistent radio emitting relativistic jets (Paredes
et~al. \cite{Paredes02}), and because of its proposed association with the unidentified source
3EG~J1824$-$1514 (Paredes et~al. \cite{Paredes00}) of the 3rd EGRET catalogue (Hartman et al.
\cite{3rdEGRETC}). Recently, Collmar (\cite{Collmar03}) reported the detection of an unidentified
$\gamma$-ray source by the COMPTEL experiment, and  \ls\ is one of the possible counterparts.  We
are interested in investigating if microquasars are able to generate the high-energy $\gamma$-ray
emission detected by EGRET\footnote{http://cossc.gsfc.nasa.gov/egret}, using a numerical model based
on IC scattering, taking into account energy losses, EC and SSC interactions. In this paper, we
focus our attention on the microquasar \ls. In Sect.~\ref{thephy} we present the physical scenario
of our model, and the model itself is described in Sect.~\ref{descr}. We apply the model to the
microquasar \ls\ in Sect.~\ref{applic}, and the obtained results are discussed in Sect.~\ref{disc}.

\section{The physical scenario} \label{thephy}

The scenario of our model consists on an X-ray binary system where the compact object has a
disk coplanar with the orbital plane and a jet not necessarily perpendicular to it. It
includes also a hot spherical region, called corona, surrounding the compact object. We assume
that the leptons dominate the radiative processes that generate $\gamma$-rays.
Relativistic electrons, already accelerated and flowing away into the jet, are exposed to
photons from the star (assumed to be an isotropic photon field) as well as to photons from the
synchrotron emission of the same population of electrons, since we suppose the presence of a
magnetic field. This magnetic field is tangled in such a way that the resulted synchrotron emission
is isotropic in the jet's reference frame. Although some features like disk-jet interaction (e.g.
for blazars, Dermer et~al. \cite{Dermer92}), reacceleration of the jet's particles (e.g. 
Markoff et~al. \cite{Markoff01}), corona-jet interaction and pair creation-annihilation
phenomena (e.g. both for Cygnus~X-1, Romero et~al. \cite{Romero02}) could be significant at
some level, their study is beyond the scope of this paper and they will be treated in upcoming
works.  
The $\gamma$-ray emitting region, the $\gamma$-jet, is assumed to be closer to the
compact object than the observed radio jets.  

\section{Description of the model} \label{descr}

We have developed a numerical IC model for the $\gamma$-ray emission from relativistic
electrons in a cylindrical jet (the $\gamma$-jet). This $\gamma$-jet is assumed to be short enough
to allow the magnetic field ($B_{\gamma}$) to be considered constant. We have included the
interaction between the relativistic electrons and both the magnetic and the radiation
fields. The energy losses of the leptonic plasma in the $\gamma$-jet are due to its own synchrotron
emission, the SSC scattering and the EC scattering of the stellar photons by the relativistic
electrons. Due to the strong losses, the electron energy distribution density
along the $\gamma$-jet model varies significantly, and in this sense the $\gamma$-jet is
non-homogeneous. Thus, the $\gamma$-jet is studied by splitting it into cylindrical transverse
cuts or slices. The size of the slices has to be suitable to get almost
homogeneous physical conditions within them (energy densities for the radiation and the
electrons).  

The code used for our $\gamma$-jet model is run in two steps. The first computes,
for each slice, both the total radiation energy densities and the radiation energy densities
per frequency unit from both the star and the synchrotron process. Then, the calculated
radiation energy densities are introduced jointly with the electron distribution function (see
Sect.~3.1) in the second step of the program to calculate the IC emission at each slice.
Because such slices evolve due to the changes of both the electron energy and the radiation
energy densities, the resulting IC radiation will be different depending on the {\it age} of
the slice (i.e. depending on the distance to the compact object). Once we have computed the IC
emission from all the slices, we can sum the contribution of all of them to get the
whole emission. Since our model is applied to a wide range of energies for both the electrons
and the seed photons, we have taken into account both the low and the high energy regimes
(i.e. the Thomson and Klein-Nishina regimes). We have used the cross section of Blumenthal \&
Gould (\cite{Blumenthal&Gould70}). The free parameters of the model are just
three: $B_{\gamma}$, the maximum electron Lorentz factor ($\gamma_{\rm e0}^{\rm
max}$) and the leptonic kinetic luminosity or leptonic jet power ($L_{\rm ke}$).

\subsection{Energy distribution of the electrons}

The differential equation that describes the electron energy evolution in our model is:
\begin{equation}\frac{dE}{dt}=-(a_{\rm s}~B^2_{\gamma}+a_{\rm c}
~U_{\rm syn}+a_{\rm c}~U_{\rm star})~E^2
\label{eq:eqdif}
\end{equation}
where a$_{\rm s}$ and a$_{\rm c}$ are constants (of values $2.37\times10^{-3}$ and
$3.97\times10^{-2}$ respectively, in cgs units), $U_{\rm star}$ is the companion star's
radiation energy density, $U_{\rm syn}$ is the synchrotron radiation 
energy density, computed using the local approximation (Ghisellini et~al. \cite{Ghisellini85}),
and $E$ is the electron energy. It is worth noting that we work now in the jet's reference
frame. Solving this equation for the given conditions, after some algebra we get the electron
Lorentz factor evolution function:
\begin{equation}
\gamma_{\rm e}=\frac{\gamma_{\rm e0}}{1+\gamma_{\rm e0}m_{\rm
e}c^2[C_{\rm s}(t-t_0)+\frac{C_{\rm c}}{v_{\rm jet}^2t'\sin
\varpi}(\arctan\frac{t-t'\cos \varpi}{t'\sin \varpi}-\arctan\frac{t_0-t'\cos
\varpi}{t'\sin \varpi})]}
\label{eq:enevol}
\end{equation}
where $t$ is the time and $t_0$ is the starting time in slice evolution. Electrons are injected
at a distance $z_0$ to the compact object, which corresponds to the time  $t_0$, and 
their evolution continues along the jet axis (to farther $z$). $v_{\rm jet}$ is the jet velocity,
$\varpi$ is the angle between the jet and the orbital plane, and $\gamma_{\rm e0}$ is the value
for the electron Lorentz factor at $t_0$. $R_{\rm orb}$ is the orbital radius of the binary
system, and $t'$ is $R_{\rm orb}/v_{\rm jet}$. $C_{\rm s}$ is $(a_{\rm s}B^2_{\gamma}+a_{\rm
c}U_{\rm syn})$, and $C_{\rm c}$ is $a_{\rm c}\Gamma_{\rm jet}L_{\rm star}/4\pi c$, where
$L_{\rm star}$ is the bolometric luminosity of the companion star, $\Gamma_{\rm jet}$ is the
Lorentz factor of the jet and $c$ is the speed of light. We describe the energy distribution of the
(injected) electrons, $N$, with a power law in the slice at t$_0$. The initial slice (like the
following ones) is assumed to be spatially homogeneous and the particle motion direction
follows an isotropic distribution (in the reference frame of the jet). So, at $t_0$:  
\begin{equation}
N(\gamma_{\rm e0})=Q\gamma_{\rm e0}^{-p},~~$for$~~\gamma_{\rm e0}^{\rm min} \le \gamma_{\rm
e0} \le \gamma_{\rm e0}^{\rm max} 
\label{eq:eldistr}
\end{equation}
where $Q$ is the normalization constant of the electrons, related to the jet power
(Georganopoulos et~al. \cite{Georganopoulos02}), and $p$ is the power law index of the
electron energy distribution function. $\gamma_{\rm e0}^{\rm max}$ is the maximum Lorentz
factor of the electrons in the first slice, which will be treated later. 
$\gamma_{\rm e0}^{\rm min}$ is the minimum electron Lorentz factor in the first slice, whose
value is determined according to both the upper limit of the seed photon energy and the lower limit
of the outgoing photon energy (see Sect.~3.3). Taking into account the conservation of the number
of particles, $N(\gamma_{\rm e},t)d\gamma_{\rm e}=N(\gamma_{\rm e0})d\gamma_{\rm e0}$, the time
evolution of Eq.~(\ref{eq:eldistr}) is found:  
\begin{equation}
N(\gamma_{\rm e},t)=Q\gamma_{\rm e}^{-p}\left[1-\gamma_{\rm e}m_{\rm
e}c^2\left(C_{\rm s}(t-t_0)+\frac{C_{\rm c}}{v_{\rm jet}^2t'\sin \varpi}
\left(\arctan\frac{t-t'\cos \varpi}{t'\sin \varpi}-\arctan\frac{t_0-t'\cos
\varpi}{t'\sin \varpi}\right) \right)\right]^{p-2}
\label{eq:evodistr}
\end{equation}

\subsection{The radiation fields}

Assuming that the stellar photon energy density is isotropic,
$U_{\rm star}$ in the reference frame of the jet is given by~~$\Gamma_{\rm jet} L_{\rm
star}/4\pi c d^2$, where $d$ is the distance between the companion star and a certain
slice of the $\gamma$-jet. Assuming that the star emits like a black-body  and taking into account
its spectral type (see 
Sect.~4.1), we can obtain the radiation energy density per frequency unit ($U_{\rm \nu~star}$).
The numerical calculation of $U_{\rm syn}$ is difficult when this seed photon field becomes the
dominant one. The problem is to know previously the evolution of the $U_{\rm syn}$ within a
slice along the $\gamma$-jet. To know $U_{\rm syn}$, $N(\gamma_{\rm e}, t)$ has to be
determined, but $N(\gamma_{\rm e},t)$ depends on the value of $U_{\rm syn}$ in the previous
slice, as can be seen in Eq.~(\ref{eq:evodistr}). We solve this in the following way: $U_{\rm
syn}$ and the radiation energy density per frequency unit ($U_{\rm \nu~syn}$) are found
initially starting with Eq.~(\ref{eq:eldistr}), determining their own values for the next step
and so on. The synchrotron emission is numerically integrated following the standard formulae
from Pacholczyk (\cite{Pacholczyk70}). The two integrals for computing numerically the emission
and absorption coefficients are: 
\begin{equation}
j_{\nu}(z)=c_3B_{\gamma}\int_{\gamma_{\rm e}^{\rm min}(z)}^{\gamma_{\rm e}^{\rm max}(z)}
N(\gamma_{\rm e},z)F(\nu/\nu_{\rm c})
d\gamma_{\rm e}
\label{eq:syem}
\end{equation}
\begin{equation}
k_{\nu}(z)=-\frac{c_3B_{\gamma}c^2}{2m_{\rm e}c^2\nu^2}\int_{\gamma_{\rm e}^{\rm min}(z)}
^{\gamma_{\rm e}^{\rm max}(z)}\frac{d}{d\gamma_{\rm e}}\left(\frac{N(\gamma_{\rm e},z)}
{\gamma_{\rm e}^2}\right)\gamma_{\rm e}(z)^2F(\nu/\nu_{\rm c})
d\gamma_{\rm e}
\label{eq:syabs}
\end{equation}
where $c_3$=$1.87\times10^{-23}$ (in cgs units) and $F(\nu/\nu_{\rm
c})=x\int_x^{\infty}k_{5/3}(z)dz$, in which $K_{5/3}(z)$ is the Bessel function of the second
kind. $\nu$ and $\nu_{\rm c}$ are the frequency and the characteristic frequency respectively.
After some algebra, $U_{\rm \nu~syn}$ is obtained.  

\subsection{IC interaction}

The code developed for reproducing the radiation-matter IC interaction operates over
files created by the first step. The seed photon energies ($\epsilon_0$) should be such
that the scattered photons had energies inside the chosen energy band following
the relationship: 
\begin{equation}
\epsilon_{\rm max}~=~\frac{4\epsilon_0\gamma_{\rm e}^2}{1+4\epsilon_0\gamma_{\rm e}}
\label{eq:emax}
\end{equation}
In the second step, both the radiation and the
matter interact throughout the cross section of the IC process (Blumenthal \& Gould
\cite{Blumenthal&Gould70}):
\begin{equation}
\sigma(x,\epsilon_0,\gamma_{\rm e})=\frac{3\sigma_{\rm
T}}{4 \epsilon_0\gamma_{\rm e}^2}~f(x)
\label{eq:scatt}
\end{equation}
where $\sigma_{\rm T}$ is the Thomson cross section constant and $f(x)$ is:
\begin{equation}
f(x)=\left[~2x\ln x+x+1-2x^2+\frac{(4\epsilon_0\gamma_{\rm
e}x)^2}{2(1+4\epsilon_0\gamma_{\rm e} x)}~\right]P(1/4\gamma_{\rm
e}^2,~1,~x)
\label{eq:fkn}
\end{equation}
in which $x$ is:
\begin{equation}
x=\frac{\epsilon}{4\epsilon_0\gamma_{\rm
e}^2(1-\epsilon/\gamma_{\rm e})} 
\label{eq:xkn}
\end{equation}
and 
\begin{equation}
P(1/4\gamma_{\rm e}^2,~1,~x)=1,~~$for$~~1/4\gamma_{\rm e}^2 \le x \le
1,~~$and$~0~$otherwise$.
\end{equation}
$P(1/4\gamma_{\rm e}^2,~1,~x)$ restricts the cross section to a physical case, where
$\epsilon$ cannot be lower than $\epsilon_0$ or higher than $\epsilon_{\rm max}$. 

The resulting IC spectral energy distribution in the observer's reference frame is:
\begin{equation}
\epsilon L_{\epsilon}=\delta^{2+p}
\epsilon^2 \sum^{z_{\rm max}}_{z_{\rm min}} 
V_{\rm slice}(z)
\int^{\epsilon_0^{\rm max}(z)}_{\epsilon_0^{\rm min}(z)}
\int^{\delta\gamma_{\rm e}^{\rm max}(z)}_{\gamma_{\rm e}^{\rm min}(z)}
U_{\rm total}(\epsilon_0,z)N(\gamma_{\rm e},z)\sigma(x,\epsilon_0,\gamma_{\rm e}) 
d\gamma
d\epsilon_0
\label{eq:lumic}
\end{equation}
Here, $\delta$ is the jet Doppler factor, viz. $\delta=\left[\Gamma_{\rm
jet}~(1-\beta\cos\theta)\right]^{-1}$ and $\beta=v_{\rm jet}/c$. $V_{\rm slice}$ is a
slice volume, which varies depending on the accuracy needed by each slice, $\theta$ is the
angle between the jet and the observer line of sight, and $U_{\rm total}(\epsilon_0,z)=U_{\rm
syn}(\epsilon_0,z)+U_{\rm star}(\epsilon_0,z)$, which is the total radiation energy density per
seed photon energy unit and slice, taking into account both the synchrotron and the star
radiation field components.  The integral has been calculated numerically taking suitable
integration steps from the behavior of the involved functions. We note that this is the
optically thin case because, in the $\gamma$-jet conditions  and at the EGRET energy range, the
IC absorption coefficient is negligible. It is worth mentioning that the electron energy
evolves following Eq.~(\ref{eq:enevol}). Therefore, in order to obtain the desired high-energy
photons, the frequency range for the seed photons will change with the evolution of the
electron energies, since there is a link (the limits of  Eq.~(\ref{eq:xkn}) induced by the 
function $P(1/4\gamma_{\rm e}^2,~1,~x)$) between the incoming photon energy, the outgoing
photon energy and the scattering electron energy. The electron Lorentz factor
integration limits in Eq.~(\ref{eq:lumic}) have been adopted following Georganopoulos et~al.
(\cite{Georganopoulos01}).

\section{Application of the model} \label{applic}

\subsection{Parameter choice for \ls} 

Although this model can be applied to other cases, we will test its  prediction
capabilities with \ls. The optical counterpart of \ls\ is a bright (V~$\sim$~11) star of
ON6.5~V((f)) spectral type (McSwain et~al. \cite{Mcswain04}), at an estimated distance of
2.9~kpc (Rib\'o et~al. \cite{Ribo02}). The  bolometric luminosity of the companion star
has been taken to be $L_{\rm star} \sim1.2\times10^{39}$~erg~s$^{-1}$. A new value of the
orbital period, $P=4.4267 \pm 0.0005$ days, the eccentricity, $e=0.48 \pm 0.06$, and the
semi-major axis of the orbit, $a=2.6\times10^{12}$~cm ($R_{\rm orb}=a$ except where
otherwise stated), have been determined recently by McSwain et~al. (\cite{Mcswain04}). We
adopt these values in our study. At radio wavelengths (1.5--15 GHz), the spectral index
was determined to be $\alpha\sim -0.5$  (where the flux density is $F_{\nu}\propto
\nu^{\alpha}$) by Mart{\'{\i}} et~al. (\cite{Marti98}). From this spectral index, we have
deduced the electron power law index ($p$) following the known relationship
$p=1-2\alpha=2$. From VLBI observations (Paredes et~al. \cite{Paredes00},
\cite{Paredes02}), the radio constraint $v_{\rm jet}\cos\theta\sim0.15c$ was found. The
observed spectrum over 100~MeV has been obtained from the third EGRET catalogue (Hartman
et~al. \cite{3rdEGRETC}). The total luminosity in the range 100--1500~MeV is
$2\times10^{35}$~erg~s$^{-1}$. The spectral energy index, $\eta=0.2\pm0.2$, corresponds
to an observed photon index of $2.2\pm0.2$ (photon energy distribution
$\phi_{\epsilon}\propto \epsilon^{-2.2}$). All these parameters are summarized in
Table~\ref{knopar}.

\begin{table*}\caption[]{Known parameters for \ls.}
\begin{flushleft}\begin{tabular}{l c c c c
c}\noalign{\smallskip} \hline \noalign{\smallskip} Parameter &
Description & Value \cr\noalign{\smallskip} \hline
\noalign{\smallskip} $a$ & orbital semi-major axis &
$2.6\times10^{12}$~cm \cr $L_{\rm star}$ & star total luminosity &
$1.2\times10^{39}$~erg~s$^{-1}$ \cr $\beta\cos\theta$ & relationship
between the jet velocity and the jet-observer line of sight angle & 0.15 \cr $L_{\rm
EGRET}$ & luminosity at the EGRET band & $2\times10^{35}$~erg~s$^{-1}$ \cr
$\eta$ & spectral energy index at the band 100--1000~MeV & 0.2 \cr
$\alpha$ & spectral index at the band 1.5--15~GHz & $-0.5$
\cr\noalign{\smallskip} \hline\end{tabular}
\end{flushleft}
\label{knopar}
\end{table*}

McSwain \& Gies (\cite{Mcswain02}) proposed a value of  $\theta=30^{\circ}$. Having this, and
the radio constraint shown above ($\beta\cos\theta\sim0.15$), we obtain a $v_{\rm jet}/c=0.2$
($\Gamma_{\rm jet}=1.02$). We stress that our model is weakly sensitive to changes on $v_{\rm
jet}$ and $\theta$ around the quoted values because of the adopted low value of the $\Gamma_{\rm
jet}$. Such a low $\Gamma_{\rm jet}$ is not in disagreement with the value found in
similar systems (e.g. 1.04 for SS~433, Spencer \cite{Spencer79}, Hjellming \& Johnston
\cite{Hjellming&johnston81}). For  \ls, we have assumed  $\varpi=90^{\circ}$. In order to fix
the $\gamma$-jet radius, we have imposed that $R_{\gamma}$ should be neither wider than the
inner disk (a few $10^7$~cm) nor thinner than some few electron Larmor radii ($\sim 10^6$~cm).
Thus, $R_{\gamma}$ has been taken to be $10^7$~cm. Regarding $z_0$, since in this scenario
($R_{\rm orb}\gg z_0$) its size does not affect the results, it is taken to be zero. To
determine the power of the jet, we need to fix also the accretion luminosity of the disk
(Falcke \&  Biermann \cite{Falcke&Biermann96}). A typical value for a microquasar with an O-type
stellar companion can be $L_{\rm ac}\simeq 10^{-8}~M_{\odot}c^2$~yr$^{-1}$. All these
parameters are summarized in Table~\ref{fixval}. 

\begin{table*}\caption[]{Fixed
parameters used in our model for \ls.}
\begin{flushleft}\begin{tabular}{l c c c c
c}\noalign{\smallskip} \hline \noalign{\smallskip} Parameter &
Description & Adopted value \cr\noalign{\smallskip} \hline
\noalign{\smallskip} $\varpi$ & angle between the jet and the orbital plane &
$90^{\circ}$ \cr $\Gamma_{\rm jet}$ & jet Lorentz factor & 1.02 \cr
$\theta$ & angle between the jet and the observer line of sight & $30^{\circ}$
\cr $R_{\gamma}$ & $\gamma$-jet radius & $10^7$~cm \cr $L_{\rm ac}$ 
& accretion disk luminosity & $10^{-8}~M_{\odot}c^2$~yr$^{-1}$ \cr\noalign{\smallskip}
\hline\end{tabular}
\end{flushleft}
\label{fixval}
\end{table*}

To determine $L_{\rm ke}$, we will use the jet-disk coupling hypothesis for
Galactic jet sources of Falcke \&  Biermann (\cite{Falcke&Biermann96}): $L_{\rm
k}\sim0.1$--$0.001L_{\rm ac}$, where $L_{\rm k}$ is the total kinetic luminosity of the jet.
Since not all the jet power is carried just by the electron population, the former scaling
relationship turns to $L_{\rm ke}\propto L_{\rm ac}$, where the prior scaling factor limits are now
upper limits. In fact, $L_{\rm ke}$ will be fixed through comparison between the observed fluxes
and the model. From the EGRET energy range, and the involved photon and electron energy
relationships shown above, the initial maximum Lorentz factor of the electrons should be about
$10^4$. A more accurate value will be found when we try to reproduce the observed spectrum slope.
Regarding $B_{\gamma}$, we will study our model behavior along a wide range of 
magnetic values, from $B_{\gamma}$ close to estimated radio jet upper values to 100~G.

\subsection{Results}

First, we explore the range of validity of the free parameters of the model taking into
account observational constraints. Once this is done, we can reproduce the observational data
with our model. Finally, additional comments on the model are made. 

\subsubsection{Exploring $B_{\gamma}$ and $\gamma_{\rm e0}^{\rm max}$}

\begin{figure}
\resizebox{\hsize}{!}{\includegraphics{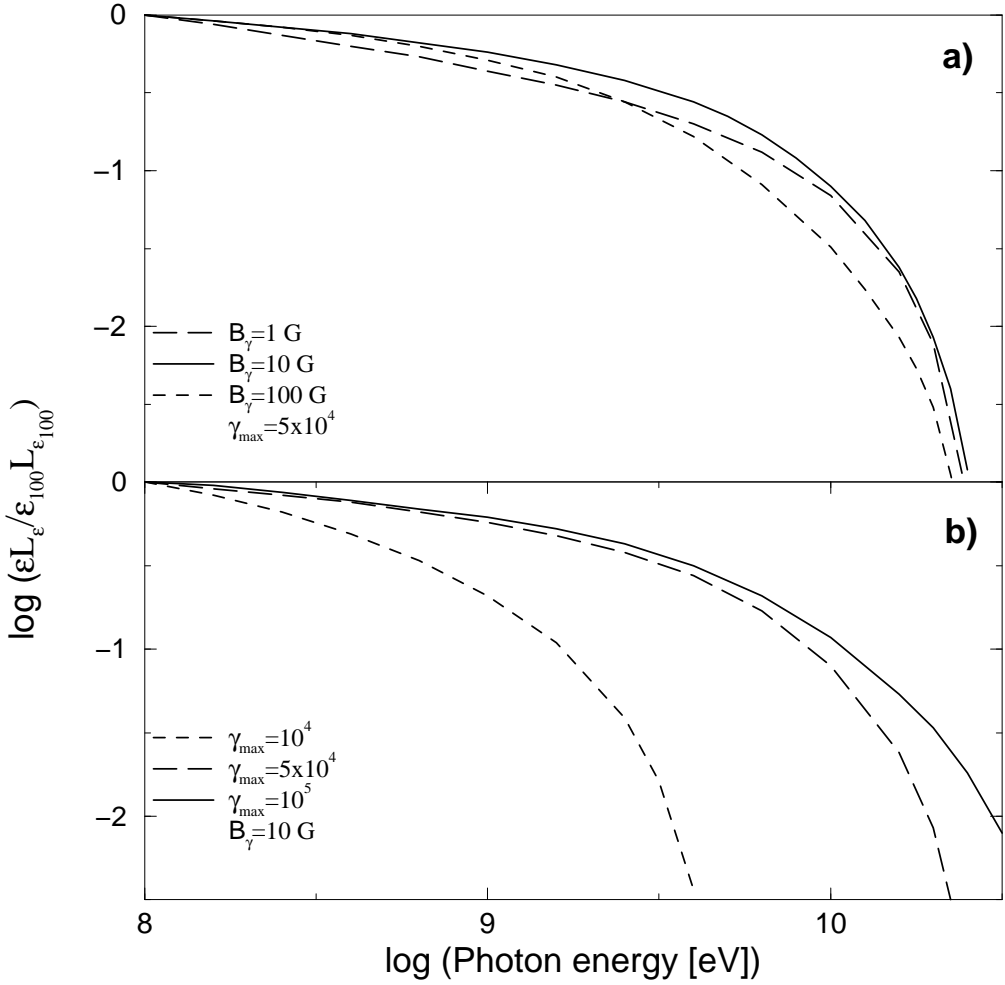}} \caption{Computed IC spectral energy
distributions normalized to their value at 100~MeV for different values of the
magnetic field strength and the maximum electron Lorentz factor. \textbf{a)} For values of the
magnetic field of 1, 10 and 100~G and fixing the maximum Lorentz factor at $5\times10^4$.
\textbf{b)} For values of the maximum Lorentz factor of the electrons of $1\times10^4$, 
$5\times10^4$ and $1\times10^5$ and fixing the magnetic field to 10~G.}  
\label{figcases}
\end{figure}

To determine the behavior of the model with magnetic field strength, we have run our model for
different values of the magnetic field strength and a fixed value of $\gamma_{\rm e0}^{\rm
max}=5\times10^4$. We show at the top of Fig.~\ref{figcases} the computed IC spectral energy
distribution ($\epsilon L_{\epsilon}$) for three representative cases (1, 10 and 100~G). The
spectral slopes in the energy range 100--1000~MeV, where EGRET detected the source, are  similar.
Since our model generates similar spectra for a wide range of magnetic field values, we have to
focus our attention on the distances reached by the $\gamma$-jet, which is strongly dependent on
$B_{\gamma}$. Then, high magnetic fields (tens of G or higher) would imply a $\gamma$-jet evolving
very fast, whereas low magnetic fields (few G or lower) would imply a $\gamma$-jet evolving smoothly
(see Sect.~4.2.3, Figs.~\ref{figdens1}~and~\ref{figdens2}). Moreover, regarding values well below
1~G, we notice that upper limits of 0.2~G  have been estimated from radio observations (Paredes
et~al. \cite{Paredes02}) for the magnetic field in the radio jet. Therefore, $B_{\gamma}$ of about
several tenths of G could be taken as a lower limit.  

In order to explore the dependence of the IC spectral energy distribution on $\gamma_{\rm
e0}^{\rm max}$, three $\gamma_{\rm e0}^{\rm max}$ ($10^4$, $5\times10^4$ and
$10^5$) have been used, fixing $B_{\gamma}=10$~G. The results are shown at the bottom of
Fig.~\ref{figcases}. For $\gamma_{\rm e0}^{\rm max}=10^4$, the computed
$\eta$, about 0.7, is 2.5~$\sigma$ softer than that given by EGRET observations.
Therefore, $\gamma_{\rm e0}^{\rm max}>10^4$ is needed. An upper limit cannot be clearly fixed
yet, since we do not know properly the spectral energy cutoff.

\subsubsection{Reproducing the EGRET data}

\begin{figure}
\resizebox{\hsize}{!}{\includegraphics{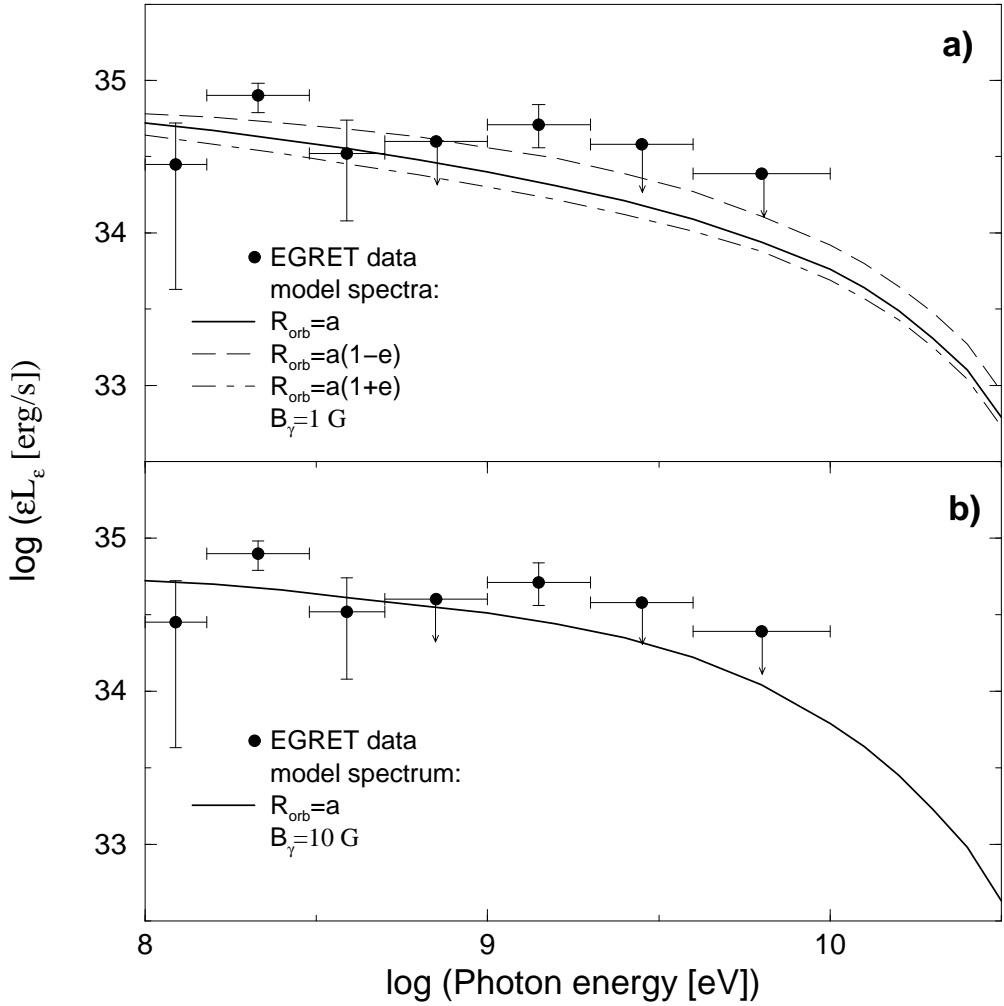}} \caption{Two IC spectral energy
distributions computed with the present model using the physical parameters of
Table~\ref{parval}. The EGRET data points are also shown. The upper limits on
undetected EGRET points are plotted with arrows. \textbf{a)} A magnetic field of 1~G has
been assumed. Also, the IC spectral energy distribution for both the apastron and the periastron
passage are shown. \textbf{b)} A magnetic field of 10~G has been assumed.}  
\label{figmodel}
\end{figure}

The computed IC spectral energy distributions for two different values of the magnetic field
are shown in Fig.~\ref{figmodel}. We also show the EGRET data points of 3EG~J1824$-$1514
(Hartman et~al. \cite{3rdEGRETC}). The kinetic luminosity has been adopted such that
the level of the observed flux at 100~MeV can be reproduced. The parameter values used in both
cases are summarized in Table~\ref{parval}. Since we are dealing with an eccentric system, the
star's radiation density in the jet is affected by the orbital distance variation. We have
explored the importance of this effect by computing the IC spectral energy distribution at
periastron and apastron passage as well as for $R_{\rm orb}=a$. In the case $B_{\gamma}=10$~G,
since SSC is strongly dominant, the effect of the orbital eccentricity on the computed IC
luminosity is not significant. However, for a $B_{\gamma}=1$~G, the EC scattering is dominant
and the orbital distance variation between apastron and periastron passages produces a change
of about 50 per cent in the computed IC luminosity. At the top of Fig.~\ref{figmodel}, we
have plotted our results obtained at three orbital distances: $1.35\times10^{12}$~cm for
periastron, $2.6\times10^{12}$~cm for the semi-major axis of the orbit and
$3.85\times10^{12}$~cm for apastron. Since the viewing periods of EGRET are longer than
the orbital period of the system, we are not able to relate the variability predictions of
our model to the variability detected by EGRET. As can be seen in Fig.~\ref{figmodel}, our
model reproduces properly the EGRET data below about 1~GeV. Beyond this energy, the computed
spectrum becomes significantly softer, but it is not in disagreement with the upper limits
given by EGRET at this energy range. 

\begin{table*}\caption[]{Parameter values used in the model for Fig.~\ref{figmodel}.}
\begin{flushleft}\begin{tabular}{l c c c c
c}\noalign{\smallskip} \hline \noalign{\smallskip} Parameter & top
& bottom \cr\noalign{\smallskip} \hline 
\noalign{\smallskip} $B_{\gamma}$ & 1~G & 10~G \cr $L_{\rm ke}$ &
$10^{36}$~erg/s & $3\times10^{36}$~erg/s \cr $\gamma_{\rm e0}^{\rm max}$ &
$10^5$ & $10^5$ \cr\noalign{\smallskip}
\hline\end{tabular} 
\end{flushleft}
\label{parval}
\end{table*}

\subsubsection{Further comments}

\begin{figure}
\resizebox{\hsize}{!}{\includegraphics{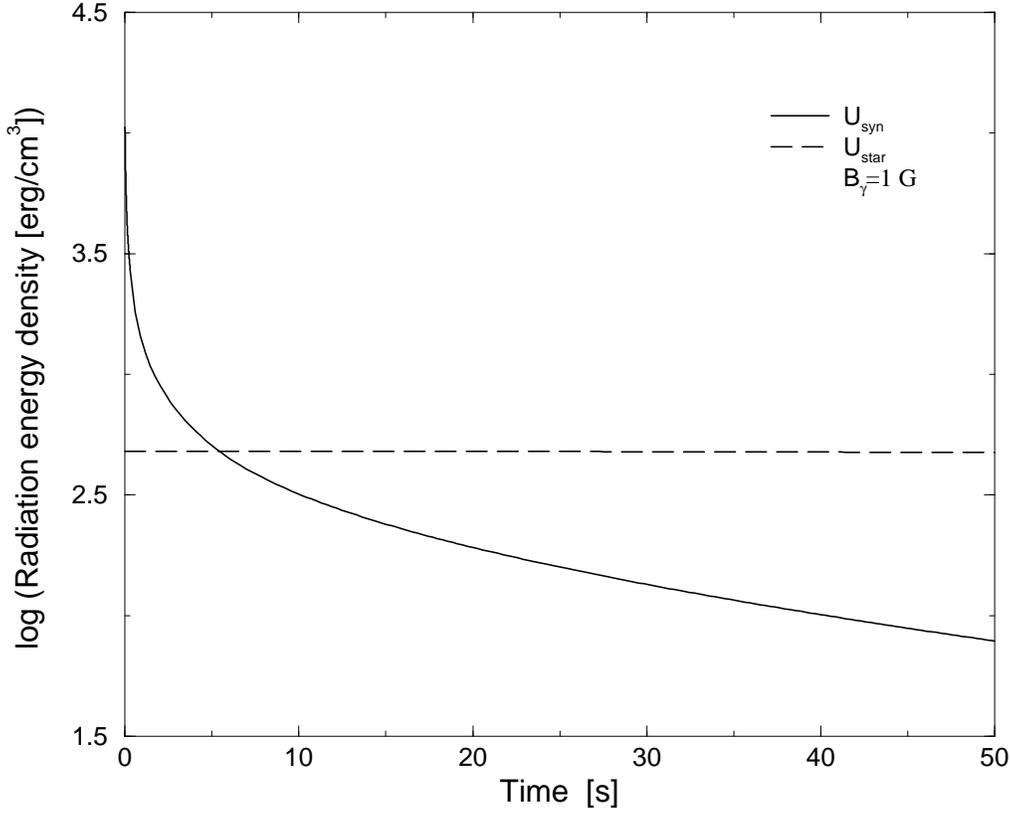}} \caption{The evolution of the synchrotron
and companion star radiation energy densities within a given slice. The value of the
magnetic field is $B_{\gamma}=1$~G and the maximum Lorentz factor of the electrons is
$5\times10^4$. The time axis finishes when a given slice moving along the jet stops emitting
due to losses through the IC effect at 100~MeV.}    
\label{figdens1}
\end{figure}

\begin{figure}
\resizebox{\hsize}{!}{\includegraphics{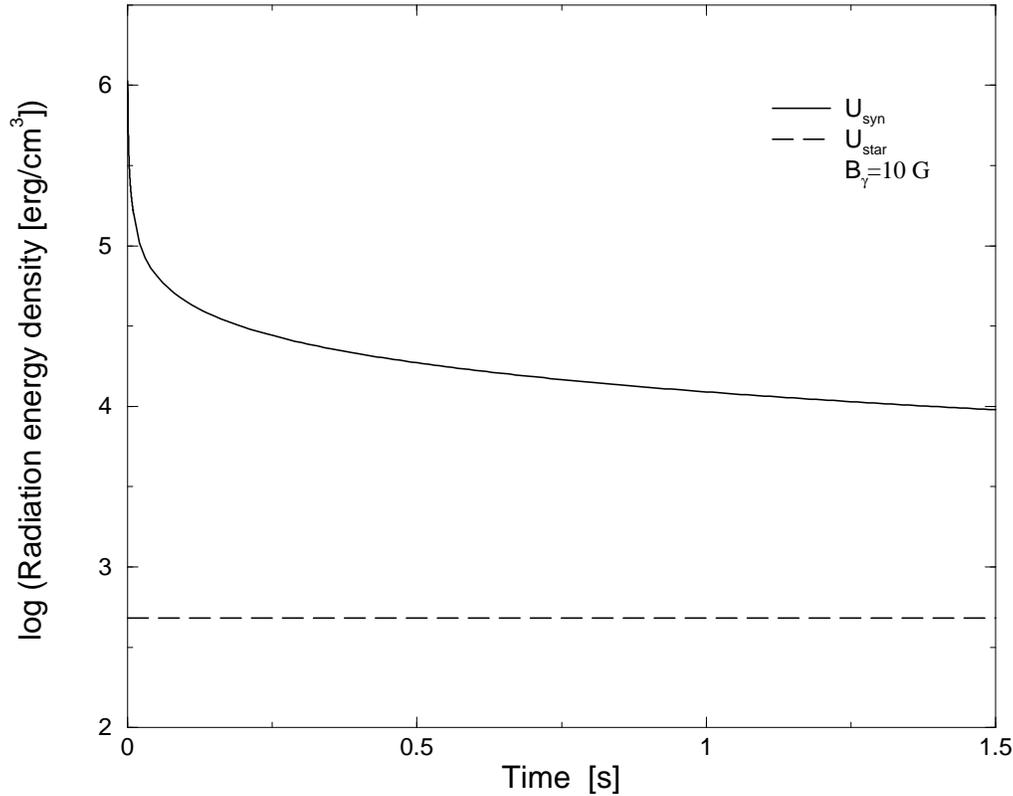}} \caption{Same as Fig.~\ref{figdens1} but
for $B_{\gamma}=10$~G.} 
\label{figdens2}
\end{figure}

Additional comments have to be made about the magnetic field in the $\gamma$-jet. We have plotted
the temporal evolution of the synchrotron and companion star radiation energy densities for both a
$B_{\gamma}$ of 1~G and a $B_{\gamma}$ of 10~G within a given slice moving along the jet
(Figs.~\ref{figdens1}~and~\ref{figdens2}). It shows that a variation of one order of magnitude in
$B_{\gamma}$ implies a variation of more than two orders of magnitude in the synchrotron radiation
energy density. Therefore, in the context of our model, magnetic fields between 1 and 10~G or higher
could be tightly related to the high-energy $\gamma$-ray emission in \ls, as the SSC loss process is
important and even dominant. Finally, we have explored the infrared synchrotron emission from the
$\gamma$-jet. Taking a $B_{\gamma}=10$~G, we have obtained a total luminosity in the range
10$^{12}$--10$^{14}$~Hz of about $10^{32}$~erg~s$^{-1}$ for \ls. If we compare this value to the
luminosity of the companion star in this frequency band, we find that the infrared jet luminosity
is five orders of magnitude lower. 

\section{Discussion} \label{disc}

As we have shown, our model is able to reproduce the high-energy $\gamma$-ray emission of the
EGRET source 3EG~J1824$-$1514 by assuming its origin in the microquasar \ls. Since there are no
special constraints associated with a particular source in our model, it could be applied to
other unidentified EGRET sources that might be also associated with microquasars. A maximum
electron Lorentz factor higher than $10^4$ is a good parameter range to obtain the
observed spectral slope. An upper limit on $\gamma_{\rm e0}^{\rm max}$ has not been determined
yet due to the lack of knowledge about the high-energy cutoff. Further observations by the
next, more sensitive, hard $\gamma$-ray instruments
(AGILE\footnote{http://agile.mi.iasf.cnr.it}, GLAST\footnote{http://glast.gsfc.nasa.gov}) are
needed to determine this cutoff. To reproduce the observed luminosities, $L_{\rm ke}\sim
10^{36}$~erg/s is needed for a magnetic field of 1~G, and $L_{\rm ke}\sim
3\times10^{36}$~erg/s is needed for a magnetic field of 10~G. Both values for $L_{\rm ke}$ are
well within the range of allowed values established in the parameter discussion.

For \ls, regarding the intensity of the magnetic field, the SSC effect would be dominant for
values of about 10~G or higher, and at least non-negligible for values of about 1~G. In the
second case,  due to the importance of the star's seed photons, variations in the orbital
distance have a significant effect on the IC flux on the scale of the orbital period of the system.  
In the case of high $B_{\gamma}$ (above 10~G), due to the strong energy losses, the electrons
might need to be reaccelerated after leaving the $\gamma$-jet to reach the observed radio jet.
However, for $B_{\gamma}=1$--$10$~G, reacceleration might not be necessary in regions closer 
to the compact object than the radio jet.

Our approach is an attempt to join radiation and matter in a complete physical way.  The
introduction of losses is a necessary step to obtain realistic results, and a way to determine at
the end the importance that these losses can have. The importance of the synchrotron process gives
strong relevance to the magnetic field. Such a question is relevant studying the seed photon origin
in the IC interaction and the timescales of the energy processes involved, and it is also related to
variability. Additional elements in our scenario such as the corona, disk, environment interaction,
reacceleration processes, pair creation-annihilation phenomena and accretion variability will be 
considered in future studies. Also, emission at other wavelengths will require a proper scenario,
different from the one presented here.

Although future observations have to be done to confirm the nature of \ls\ as a $\gamma$-ray
emitter, our numerical model is able to produce significant $\gamma$-ray emission levels. AGILE, an
italian hard $\gamma$-ray satellite expected to be in orbit in 2005, with an angular resolution two
times better than EGRET, will make it possible to improve the accuracy of the hard $\gamma$-ray
source position. Also, its sensitivity, almost three times better than that of EGRET, could
define more clearly the spectral shape and its cutoff. The next generation of hard $\gamma$-ray
satellites represented by GLAST, which will be put in orbit around 2008, will improve even more the
study of the  variability and spectral features at this energy range, in order to better understand
the physics underlying the very high energy processes. 

\begin{acknowledgements}

We are grateful to Gustavo Romero, Marc Rib\'o, and Joan Garc{\'{\i}}a-S\'anchez for their
useful comments and suggestions.V.B-R. and J.M.P. acknowledge partial support by DGI of the
Ministerio de Ciencia y Tecnolog{\'{\i}}a (Spain) under grant AYA-2001-3092, as well as
additional support from the European Regional Development Fund (ERDF/FEDER). During this work,
V.B-R has been supported by the DGI of the Ministerio de Ciencia y Tecnolog{\'{\i}}a (Spain)
under the fellowship FP-2001-2699.  

\end{acknowledgements}

{}


\begin{thebibliography}{}

\bibitem[1999]{Atoyan&aharonian99} 
Atoyan, A.~M.~\& Aharonian, F.~A.\ 1999, \mnras, 302, 253 

\bibitem[1986]{Band&grindlay86}
Band, D.~L.~\& Grindlay, J.~E.\ 1986, \apj, 311, 595 

\bibitem[1970]{Blumenthal&Gould70}
Blumenthal, G. R.~\& Gould, R. J. 1970, RMP, 42, 237 

\bibitem[2003]{Collmar03}
Collmar, W. 2003, Proc. 4th AGILE Science Workshop, Frascati (Rome) on 11--13 June 2003

\bibitem[1992]{Dermer92}
Dermer, C.~D., Schlickeiser, R., \& Mastichiadis, A.\ 1992, \aap, 256, L27

\bibitem[1996]{Falcke&Biermann96}
Falcke, H.~\& Biermann, P.~L. 1996, A\&A, 308, 321 

\bibitem[2004]{Fender04}
Fender, R., to appear in 2004, Compact Stellar X-Ray Sources, eds. W.H.G. Lewin and
M. van der Klis, Cambridge University Press

\bibitem[2001]{Georganopoulos01}
Georganopoulos, M., Kirk, J.~G., \& Mastichiadis, A.\ 2001, \apj, 561, 111 

\bibitem[2002]{Georganopoulos02}
Georganopoulos, M., Aharonian, F.~A., \& Kirk, J.~G. 2002, A\&A, 388, L25 

\bibitem[1985]{Ghisellini85}
Ghisellini, G., Maraschi, L., \& Treves, A. 1985, A\&A, 146, 204 

\bibitem[1999]{3rdEGRETC}
Hartman, R. C., Bertsch, D.~L., \& Bloom, S.~D.~et al. 1999, \apjs, 123, 79

\bibitem[1981]{Hjellming&johnston81} 
Hjellming, R.~M.~\& Johnston, K.~J.\ 1981, \apjl, 246, L141 

\bibitem[2002]{Kaufman-Bernado02}
Kaufman Bernad\'o, M.~M., Romero, G.~E., \& Mirabel, I.~F. 2002, A\&A, 385,
L10--L13 

\bibitem[2001]{Markoff01} 
Markoff, S., Falcke, H., \& Fender, R.\ 2001, \aap, 372, L25 

\bibitem[1998]{Marti98}
Mart{\'{\i}}, J., Paredes, J.~M., \& Rib\'o, M. 1998, A\&A, 338, L71

\bibitem[2002]{Mcswain02}
McSwain, M.~V.~\& Gies, D.~R. 2002, ApJ, 568, L27

\bibitem[2004]{Mcswain04}
McSwain, M.~V., Gies D.~R., Huang W.,~et al. 2004, ApJ, 600, 927

\bibitem[1999]{Mirabel&rodriguez99}
Mirabel, I.~F.~\& Rodr{\'{\i}}guez, L.~F. 1999, ARA\&A, 37, 409 

\bibitem[1970]{Pacholczyk70}
Pacholczyk, A.~G., 1970, Radio Astrophysics, Freeman, San Francisco, CA 

\bibitem[2000]{Paredes00}
Paredes, J.~M., Mart{\'{\i}}, J., Rib\'o, M., \& Massi, M. 2000, Science, 288, 2340

\bibitem[2002]{Paredes02}
Paredes, J.~M., Rib\'o, M., Ros, E., Mart{\'{\i}}, J., \& Massi,
M. 2002, A\&A, 393, L99

\bibitem[2002]{Ribo02}
Rib\'o, M., Paredes, J.~M., \& Romero, G.~E., et~al. 2002, A\&A,
384, 954

\bibitem[2002]{Ribo02t}
Rib\'o, M. 2002, PhD Thesis, Universitat de Barcelona

\bibitem[2002]{Romero02}
Romero, G.~E., Kaufman Bernad\'o, M.~M., \& Mirabel, I.~F. 2002, A\&A, 393, L61

\bibitem[2003]{Romero03}
Romero, G.~E., Torres, D.~F., Kaufman  Bernad\'o, M.~M., \& Mirabel, I.~F. 2003, A\&A, 410, L1  

\bibitem[1979]{Spencer79} 
Spencer, R.~E.\ 1979, \nat, 282, 483 

\end{thebibliography}
\end{document}